\documentclass[prd,aps,a4paper,twocolumn,eqsecnum,nofootinbib,floatfix]{revtex4}  

\newif\ifusesec
\usesectrue  
   
\usepackage{graphicx} 
\usepackage{amsmath,amsfonts,amssymb}

\newcommand{\beq}{\begin{equation}}
\newcommand{\eeq}{\end{equation}}
\newcommand{\bea}{\begin{eqnarray}}
\newcommand{\eea}{\end{eqnarray}}
\begin{document}

\title{Momentum recoil in the relativistic two-body problem: higher-order tails}

\author{Donato Bini$^{1,2}$, Andrea Geralico$^1$}
  \affiliation{
$^1$Istituto per le Applicazioni del Calcolo ``M. Picone,'' CNR, I-00185 Rome, Italy\\
$^2$INFN, Sezione di Roma Tre, I-00146 Rome, Italy\\
}

\date{\today}

\begin{abstract}
In the description of the relativistic two-body interaction, together with the effects of energy and angular momentum losses due to the emission of gravitational radiation, one has to take into account also the loss of linear momentum, which is responsible for the recoil of the center-of-mass of the system.
We compute higher-order tail (i.e., tail-of-tail and tail-squared) contributions to the linear momentum flux for a nonspinning binary system either along hyperboliclike or ellipticlike orbits.
The corresponding orbital averages are evaluated at their leading post-Newtonian approximation, using harmonic coordinates and working in the Fourier domain.
The final expressions are given in a large-eccentricity (or large-angular momentum) expansion along hyperboliclike orbits and in a small-eccentricity expansion along ellipticlike orbits.
We thus complete a previous analysis focusing on both energy and angular momentum losses [Phys. Rev. D \textbf{104}, no.10, 104020 (2021)], providing brick-type results which will be useful, e.g., in the high-accurate determination of the radiated impulses of the two bodies undergoing a scattering process.
\end{abstract}

\maketitle

\section{Introduction}

During the coalescence process of a binary system a certain amount of the total linear momentum is carried away via gravitational wave emission and, consequently, the center-of-mass of the binary recoils, i.e., moves in order to balance the linear momentum loss.
This effect can be ascribed to the mass asymmetry between the two bodies (and also to either unequal or misaligned spins in the case of spinning binaries), and has important implications in many astrophysical scenarios, especially for systems whose host structure has escape speeds comparable with the recoil velocity (see, e.g., Ref. \cite{Schnittman:2007ij} and references therein).
The recoil, or kick, is also responsible for the eventual ejection of black holes from the host galaxy as galaxies merge.

The first explicit calculation of the linear momentum loss and associated recoil velocity within a Post-Newtonian (PN) framework is due to Fitchett~\cite{Fitchett:1983} for an inspiralling binary system of two point masses in Keplerian orbit, based on earlier works by Peres \cite{Peres:1962zz} and Bekenstein \cite{Bekenstein:1973zz}.
The 1PN corrections to this result were successively computed by Wiseman \cite{Wiseman:1992dv} and by Junker and Schafer~\cite{Junker:1992kle}, who considered also the case of hyperboliclike orbits.
The 2PN level of accuracy was then achieved by Blanchet, Qusailah and Will \cite{Blanchet:2005rj}, who limited, however, their analysis to the quasi-circular case.
Spin-orbit and spin-spin corrections were later included by Kidder \cite{Kidder:1995zr} and by Racine, Buonanno and Kidder~\cite{Racine:2008kj}, respectively, who obtained 2PN accurate expressions for the linear momentum flux valid for any kind of orbits, but then explicit results were still specialized to the quasi-circular case.
Recently, Cho, Porto and Yang \cite{Cho:2022syn} have reobtained and extended these results in the spin-spin sector at higher PN level following an effective field theory (EFT) approach \cite{Cho:2021mqw} (see also Refs. \cite{Maia:2017gxn,Maia:2017yok} for the computation of both spin-orbit and spin-spin radiation reaction effects).

Analytical or semi-analytical methods, including perturbation theory and effective-one-body approach (see, e.g., Refs. \cite{Favata:2004wz} and \cite{Damour:2006tr}), are in general unable to provide a sufficiently accurate estimate of the recoil velocity, since for coalescing binaries most of the linear momentum flux is emitted during the merger and ringdown phases, whereas radiation from hyperbolic encounters is plunge-dominated. 
Therefore, one must rely on numerical relativity simulations to make predictions for the kick velocity of the post-merger black hole,  as well as for the final distribution of the radiated momentum due to the recoil over the two individual black hole at the end of a scattering process (see, e.g., Refs. \cite{Gonzalez:2006md,Healy:2008js}).
Nevertheless, pushing PN-based computations to higher orders will allow for a more and more accurate description of the linear momentum loss as well as a reliable estimate of the recoil velocity accumulated during the inspiral phase.
In addition, the knowledge of radiative losses of energy, angular momentum and linear momentum with high-PN accuracy is necessary to evaluate the radiation-reaction contributions to physical observables, like the scattering angle in hyperbolic encounters.

After many years of works on ellipticlike orbits, a new interest in hyperboliclike motion has been recently raised, since the evaluation of the scattering angle has proven to contain all the necessary structural information to characterize the local 5PN and 6PN (conservative) Hamiltonian of the system \cite{Bini:2019nra,Bini:2020wpo,Bini:2020nsb,Bini:2020hmy,Bini:2020rzn}.
On the other side, EFT succeeded in computing the Post-Minkowskian (PM) Hamiltonian of the system at the 3PM  (complete) and 4PM (partial) levels by using various complementary methods, including the classical worldline approach and a novel technique termed double-copy, which -roughly speaking- looks at gravity as the square of a Yang-Mills theory, establishing a connection
between scattering amplitudes and classical dynamics (see, e.g., Refs. \cite{Bjerrum-Bohr:2018xdl,Cheung:2018wkq,Bern:2019nnu,Bern:2019crd,Damour:2019lcq,Kalin:2020mvi,Kalin:2020fhe,Bern:2021dqo,Dlapa:2021npj,Bern:2021yeh,Dlapa:2021vgp}).
It has been known for a long time that the radiation-reaction linear momentum loss contributes to the total change of 4-momentum of each body undergoing the scattering process starting at 3PM order \cite{Kovacs:1977uw,Kovacs:1978eu}.
However, the computation of the radiated impulse has been completed only very recently by using different approaches \cite{DiVecchia:2020ymx,Damour:2020tta,DiVecchia:2021ndb,Herrmann:2021tct,Bjerrum-Bohr:2021din,Bini:2021gat}.
Going to higher PM orders requires more accurate expressions for the radiative losses, which are known at their lowest PM expansion only \cite{Damour:2020tta,Herrmann:2021tct}. Fractionally 2PN-accurate expansions of the higher PM energy, angular momentum and linear momentum losses up to 7PM have been computed in Ref. \cite{Bini:2021gat}, leading to the determination of the 4PM and 5PM contributions to the radiated impulses with absolute 4.5PN accuracy.

The linear momentum flux can be written as a superposition of couplings between different radiative multipole moments \cite{Thorne:1980ru}, which can be  in turn  expressed in terms of the source multipole moments through the multipolar-post-Minkowskian (MPM) formalism \cite{Blanchet:1985sp,Blanchet:1989ki,Damour:1990ji,Blanchet:1998in,Poujade:2001ie}.
The radiative moments contain both \lq\lq instantaneous'' terms, which depend on the source moments evaluated at the retarded time, and  \lq\lq hereditary'' terms, which are instead given by \lq\lq tail'' integrals over the full past history of the source \cite{Blanchet:1987wq,Blanchet:1992br}.
The latter can be further decomposed as tail (quadratic in $G$), tail-of-tail and tail-squared (cubic in $G$), and higher nonlinear interaction terms \cite{Blanchet:1993ec,Blanchet:1997jj}.
As a result, the flux will consist of an instantaneous (local in time) part and a (nonlocal) hereditary part. 
Tail terms up to the 2.5PN level have been computed in Ref. \cite{Mishra:2011qz} for quasi-circular motion, extending previous results valid at the leading 1.5PN order \cite{Blanchet:2005rj,Racine:2008kj}.
For hyperboliclike orbits leading order tails have been instead recently computed in Ref. \cite{Bini:2021gat}.

In the present paper we evaluate higher-order tail contributions to the linear momentum flux averaged along the hyperboliclike motion at their leading PN level, completing a previous study on the analogous effects associated with both energy and angular momentum losses \cite{Bini:2021gat,Bini:2021jmj}.
We also compute both quadratic and cubic linear momentum tails for ellipticlike orbits, extending existing results out of the circular orbit limit.

Finally, let us note in passing that   we will work conveniently in the frequency domain, by using harmonic coordinates and following the notation of Ref. \cite{Bini:2021qvf}.

\section{Linear momentum tail integrals along hyperboliclike orbits}

The linear momentum flux at the leading PN order in terms of the mass-type ($U_{L}$) and current-type ($V_{L}$) radiative multipole moments (with $L=i_1i_2\cdots i_l$ being a multi-index consisting of $l$ spatial indices) is given by~\cite{Thorne:1980ru} 
\begin{eqnarray}
\label{fluxPidef}
{\mathcal{F}_{P}^{i}}(U)&\equiv&\left(\frac{d P_i}{dU}\right)^{\rm GW}\nonumber\\
&=& \frac{G}{c^7}\,\biggl[\frac{2}{63}\,
U^{(1)}_{ijk}\,U^{(1)}_{jk}+\frac{16}{45}\,\epsilon_{ijk}
U^{(1)}_{ja}\,V^{(1)}_{ka}\nonumber\\&&
+O\left({1\over c^2}\right)\biggr]\,,
\end{eqnarray}
where a superscript in parenthesis denotes repeated retarded time derivatives. 
The flux is a function of the retarded time $U=T-R/c$ in radiative coordinates, and it is related to the corresponding retarded time $u=t-r/c$ in harmonic coordinates by 
\beq
\label{Uvst}
U = t- \frac{r}{c}
-\frac{2G\mathcal{M}}{c^3}\ln\left(\frac{r}{r_0}\right) +  O \left(\frac{1}{c^5}\right)\,,
\eeq
where $\mathcal{M}$ denotes the total Arnowitt-Deser-Misner (ADM) mass of the system, and $r_0$ is a constant length scale.

Expressing the radiative multipole moments in terms of the source moments ($I_L, J_L$) \cite{Blanchet:1997jj,Blanchet:2008je} allows for decomposing the flux as the sum of instantaneous and hereditary terms,
\beq
\mathcal{F}_{P}^i(U)=\mathcal{F}^i_{P\,\rm inst}(U)+\mathcal{F}^i_{P\,\rm hered}(U)\,,
\eeq
where the instantaneous part depends on the dynamics of the system at the retarded instant $U$ only, while the hereditary part is nonlocal in time depending on the full past history~\cite{Blanchet:1987wq,Blanchet:1992br}.
The hereditary part can be further decomposed into a first-order tail part (quadratic in $G$) and a higher-order tail part (which is higher order in $G$).
We have already computed in Ref. \cite{Bini:2021gat} the first-order tail contributions to the linear momentum flux at the leading order in their PN expansion.
In the present work we will focus on cubic tails, which are referred to in the literature as tail-of tail and tail-squared terms \cite{Blanchet:1997jj}, so that the hereditary part of the flux reads
\beq
\mathcal{F}^i_{P\,\rm hered}(t)=\mathcal{F}^i_{P\,\rm tail}(t)+\mathcal{F}^i_{P\,\rm tail(tail)}(t)+\mathcal{F}^i_{P\,\rm (tail)^2}(t)\,,
\eeq
at that level of approximation, with
\begin{widetext}

\bea
\label{Ftail}
\mathcal{F}_{P\,\rm tail}^i(t) &=&\frac{G^2{\mathcal M}}{c^{10}}\left\{\frac{4}{63}\left[\,I_{ijk}^{(4)}(t)
\int_{0}^{\infty} d\tau  \ln \left({\tau\over C_{I_2}}\right) 
I^{(5)}_{jk}(t-\tau)
+\,I_{jk}^{(3)}(t)\int_{0}^{\infty} d\tau  \ln
\left({\tau\over C_{I_3}}\right) 
I^{(6)}_{ijk}(t-\tau)\right]\right.
\nonumber\\
&+&\left.
{32 \over 45}\,\epsilon_{ijk}\,\left[
I_{ja}^{(3)}(t)\int_{0}^{\infty} d\tau  \ln \left({\tau\over C_{J_2}}
\right) 
J^{(5)}_{ka}(t-\tau)
+\,J_{ka}^{(3)}(t)\int_{0}^{\infty} d \tau
 \ln \left({\tau\over C_{I_2} }\right) 
I^{(5)}_{ja}(t-\tau)\right] \right\}
\,,
\eea
and  
\bea
\label{Ftailtail} 
\mathcal{F}_\mathrm{P\, tail(tail)}^i(t) &=&\frac{G^3{\mathcal M}^2}{c^{13}}
\left\{
\frac{32}{45}\epsilon_{ijk}\left[
J^{(3)}_{ka}(t)\int_{0}^{\infty} d \tau 
\left(
\ln^2\left(\frac{\tau}{C_{I_2}}\right)-\frac{107}{105}\ln\left(\frac{\tau}{\widetilde C_{I_2}}\right) 
\right)
I^{(6)}_{ja}(t-\tau)\right.\right.\nonumber\\
&&\left.
+I^{(3)}_{ja}(t)\int_{0}^{\infty} d \tau 
\left(
\ln^2\left(\frac{\tau}{C_{J_2}}\right)-\frac{107}{105}\ln\left(\frac{\tau}{\widetilde C_{J_2}}\right) 
\right)
J^{(6)}_{ka}(t-\tau)\right]\nonumber\\
&&
+\frac{4}{63}\left[
I^{(4)}_{ijk}(t) \int_{0}^{\infty} d \tau 
\left(
\ln^2\left(\frac{\tau}{C_{I_2}}\right)-\frac{107}{105}\ln\left(\frac{\tau}{\widetilde C_{I_2}}\right) 
\right)
I^{(6)}_{jk}(t-\tau)\right.\nonumber\\
&&\left.\left.
+I^{(3)}_{jk}(t) \int_{0}^{\infty} d \tau 
\left(
\ln^2\left(\frac{\tau}{C_{I_3}}\right)-\frac{13}{21}\ln\left(\frac{\tau}{\widetilde C_{I_3}}\right) 
\right)
I^{(7)}_{ijk}(t-\tau)\right]
\right\}
\,,\nonumber\\
\mathcal{F}_\mathrm{P\, (tail)^2}^i(t)&=&\frac{G^3{\mathcal M}^2}{c^{13}}\left\{ 
\frac{8}{63}\int_0^\infty d\tau  \ln\left(\frac{\tau}{C_{I_3}}\right)  I^{(6)}_{ijk}(t-\tau)\,  
\int_0^\infty d\tau   \ln\left(\frac{\tau}{C_{I_2}}\right) I^{(5)}_{jk}(t-\tau)\right. \nonumber\\ 
&& \left.
+\frac{64}{45} \epsilon_{ijk} \int_0^\infty d\tau  \ln\left(\frac{\tau}{C_{I_2}}\right)  I^{(5)}_{ja}(t-\tau)\, \int_0^\infty d\tau  \ln\left(\frac{\tau}{C_{J_2}}\right)  J^{(5)}_{ka}(t-\tau)
\right\}
\,,
\eea
\end{widetext}
by substituting the relations (4.16), (4.17a) and (4.17b) of Ref. \cite{Blanchet:1997jj} in the general expression \eqref{fluxPidef}, PN expanding and selecting the nonlocal part only.
Here we have introduced the following set of multipolar constants ($\tau_0=cr_0$)
\bea
C_{I_2}&=& 2\tau_0 e^{-11/12}
\,,\nonumber\\
C_{I_3}&=& 2\tau_0 e^{-97/60}
\,,\nonumber\\
C_{J_2}&=& 2\tau_0 e^{-7/6}
\,,\nonumber\\
\widetilde C_{I_2}&=& C_{I_2}e^{\frac{515063}{179760}}
\,,\nonumber\\
\widetilde C_{I_3}&=& C_{I_3}e^{-\frac{18841}{109200}}
\,,\nonumber\\
\widetilde C_{J_2}&=& C_{J_2}e^{-\frac{60103}{44940}}
\,.
\eea
The quadratic-in-$G$  term \eqref{Ftail} is the dominant tail at (fractional) 1.5PN order, while the two cubic-in-$G$ order tails \eqref{Ftailtail} are both at (fractional) 3PN order.
The multipolar moment expressions agree with standard literature, see e.g., Ref. \cite{Arun:2007sg}, which we also basically follow for notation and conventions. 

The previous relations \eqref{Ftail} and \eqref{Ftailtail} can be cast in the more compact form 
\begin{widetext}
\bea
\label{Ftail_comp}
\mathcal{F}_{P\,\rm tail}^i(t) &=&\frac{G^2{\mathcal M}}{c^{10}}\left\{\frac{4}{63}\left[\,I_{ijk}^{(4)}(t)
{\mathcal T}_{\ln }[I^{(5)}_{jk};C_{I_2}](t)
+\,I_{jk}^{(3)}(t)
{\mathcal T}_{\ln }[I^{(6)}_{ijk};C_{I_3}](t)
\right]\right.
\nonumber\\
&+&\left.
{32 \over 45}\,\epsilon_{ijk}\,\left[
I_{ja}^{(3)}(t)
{\mathcal T}_{\ln }[J^{(5)}_{ka};C_{J_2}](t)
+\,J_{ka}^{(3)}(t)
{\mathcal T}_{\ln }[I^{(5)}_{ja};C_{I_2}](t)
\right] \right\}
\,,\nonumber\\ 
\mathcal{F}_\mathrm{P\, tail(tail)}^i(t) &=&\frac{G^3{\mathcal M}^2}{c^{13}}
\left\{
\frac{32}{45}\epsilon_{ijk}\left[
J^{(3)}_{ka}(t) 
\left({\mathcal T}_{\ln^2 }[I^{(6)}_{ja};C_{I_2}](t)-\frac{107}{105}{\mathcal T}_{\ln}[I^{(6)}_{ja};\widetilde C_{I_2}](t)\right)
\right.\right.\nonumber\\
&&\left.
+I^{(3)}_{ja}(t) 
\left({\mathcal T}_{\ln^2 }[J^{(6)}_{ka};C_{J_2}](t)-\frac{107}{105}{\mathcal T}_{\ln}[J^{(6)}_{ka};\widetilde C_{J_2}](t)\right)
\right]\nonumber\\
&&
+\frac{4}{63}\left[
I^{(4)}_{ijk}(t) 
\left({\mathcal T}_{\ln^2 }[I^{(6)}_{jk};C_{I_2}](t)-\frac{107}{105}{\mathcal T}_{\ln}[I^{(6)}_{jk};\widetilde C_{I_2}](t)\right)
\right.\nonumber\\
&&\left.\left.
+I^{(3)}_{jk}(t) 
\left({\mathcal T}_{\ln^2 }[I^{(7)}_{ijk};C_{I_3}](t)-\frac{13}{21}{\mathcal T}_{\ln}[I^{(7)}_{ijk};\widetilde C_{I_3}](t)\right)
\right]
\right\}
\,,\nonumber\\
\mathcal{F}_\mathrm{P\, (tail)^2}^i(t)&=&\frac{G^3{\mathcal M}^2}{c^{13}}\left\{ 
\frac{8}{63}{\mathcal T}_{\ln }[I^{(6)}_{ijk};C_{I_3}](t) {\mathcal T}_{\ln }[I^{(5)}_{jk};C_{I_2}](t)
+\frac{64}{45} \epsilon_{ijk} 
{\mathcal T}_{\ln }[I^{(5)}_{ja};C_{I_2}](t)
{\mathcal T}_{\ln }[J^{(5)}_{ka};C_{J_2}](t)
\right\}
\,,
\eea

\end{widetext}
by using the notation 
\beq
\label{typical_int}
{\mathcal T}_{\ln^m}[X^{(n)}_L;C_{X_L}](t)
=\int_{0}^\infty d\tau X^{(n)}_L(t-\tau)\ln^m \left(\frac{\tau}{C_{X_L}}\right)\,,
\eeq
introduced in Ref. \cite{Bini:2021qvf}, with $X_L$ denoting a generic multipolar moment, and $m=1,2$ are the only powers of the log terms needed here. 

These tails have been termed in Refs. \cite{Bini:2021gat,Bini:2021qvf} \lq\lq past tails,'' since they refer to the  interaction between the bodies occurring in the past, so that they are asymmetric under time-reversal.
However, one can decompose them into a time-symmetric (ts) and a time-antisymmetric (tas) part.
When dealing with ts tails it is enough to replace ${\mathcal T}_{\ln^m}\to{\mathcal T}^{\rm ts}_{\ln^m}$ in Eq. \eqref{Ftail_comp}, with
\beq
{\mathcal T}^{\rm ts}_{\ln^m}[X^{(n)}_L;C_{X_L}](t)=\int_{0}^\infty d\tau X^{(n)}_{L\, \rm sym}(t,\tau)\ln^m \left(\frac{\tau}{C_{X_L}}\right)\,, 
\eeq  
where
\beq
X^{(n)}_{L\, \rm sym}(t,\tau)=\frac12[X^{(n)}_L(t-\tau)+X^{(n)}_L(t+\tau)]\,.
\eeq

\subsection{Time-integrated loss along hyperboliclike orbits}

We will evaluate below the leading PN order contribution to the orbital averages
\bea
\label{LOtailsaver}
(\Delta P_i)_\mathrm{X}&=&\int_{-\infty}^{\infty}dt\, \mathcal{F}_\mathrm{X}^i (t)\nonumber\\
&=&(\Delta P_i)_\mathrm{X,\,ts}+(\Delta P_i)_\mathrm{X,\,tas}
\,,
\eea
along hyperboliclike orbits, where the label $X$ is either tail, tail(tail), or (tail)$^2$, and we also distinguish among time-symmetric and time-antisymmetric contributions.
Therefore, the Newtonian description of the binary dynamics suffices.
The Keplerian parametrization  of the hyperbolic motion in harmonic coordinates is
\begin{eqnarray} \label{hypQK2PN}
r&=& \bar a_r (e_r  \cosh v-1)\,,\nonumber\\
\bar n t&=& e_r  \sinh v-v\,,\nonumber\\  
\phi&=&2\, {\rm arctan}\left[\sqrt{\frac{e_r +1}{e_r -1}}\tanh \frac{v}{2}  \right]\,,
\end{eqnarray}
where we have used dimensionless variables, i.e., $r=c^2r^{\rm phys}/(GM)$, $t= c^3t^{\rm phys}/(GM)$, and the orbital parameters $\bar n$, $\bar a_r$, $e_r$ are given by
\beq
\bar n=(2\bar E)^{3/2}\,,\quad
\bar a_r=\frac1{2\bar E}\,,\quad
e_r=\sqrt{1+2\bar Ej^2}\,,
\eeq
in terms of the dimensionless energy and angular momentum parameters $\bar E$ and $j$.
The latter are defined by
\beq 
\bar E \equiv \frac{E_{\rm tot}-Mc^2}{\mu c^2}\,, \qquad
j\equiv \frac{c J}{G M \mu}\,,
\eeq
where $E_{\rm tot}$ and $J$ are the total center-of-mass energy and angular momentum of the binary system, respectively, with total mass $M=m_1+m_2$, reduced mass $\mu\equiv m_1 m_2/M$, and symmetric mass ratio $\nu={\mu}/{M}$.
At the leading order we are interested in here we can set the total ADM mass of the system $\mathcal{M}=M$. 
In addition we will set $G=M=c=1$ for simplicity.

\section{Frequency-domain computation of tail integrals}

The computation of tail integrals is more conveniently performed in the Fourier domain. 
In fact, the time-domain convolutions entering Eq. \eqref{LOtailsaver} become multiplications by the Fourier transforms of their kernels \cite{Bini:2021gat}.

\subsection{Past tails}

Let us consider past tails first.
The $m$-type past-tail \eqref{typical_int} associated with the history of $X_L^{(n)}$ becomes
\bea
\label{typical_int_n}
&&{\mathcal T}_{\ln^m}[X^{(n)}_L;C_{X_L}](t)=\nonumber\\
&&\qquad\qquad\int_{-\infty}^\infty \frac{d\omega}{2\pi}e^{-i\omega  t  }(-i\omega)^n\hat X_L(\omega) A_m (\omega, C_{X_L})
\,,\nonumber\\
\eea
upon substituting in it the Fourier expansion 
\beq
X_{L}(\tau)=\int_{-\infty}^\infty \frac{d\omega}{2\pi} \, e^{-i\omega  \tau}  \hat X_{L}(\omega) \,,
\eeq
of the generic multipolar moment $X_{L}(\tau)$.
Here
\beq
A_m (\omega, C_{X_L})=\int_{0}^\infty d\xi  e^{ i\omega  \xi  } \ln^m \left(\frac{\xi}{C_{X_L}}\right)\,,
\eeq
which for $m=1,2$ read 
\bea
A_1 (\omega, C_{X_L})&=&-\frac{\pi}{2|\omega|}-\frac{i}{|\omega|}{\rm sgn}(\omega)\ln (C_{X_L}|\omega| e^\gamma)
\,, \nonumber\\
A_2 (\omega, C_{X_L})&=& \frac{\pi}{|\omega|}\ln (C_{X_L}|\omega| e^\gamma)\nonumber\\
&+&
\frac{i}{|\omega|}{\rm sgn}(\omega)\left[\ln^2 (C_{X_L}|\omega| e^\gamma)-\frac{\pi^2}{12}  \right]
\,.\nonumber\\
\eea

Taking the orbital averages \eqref{LOtailsaver} leads to integrals of the type
\begin{widetext}
\bea
\label{basic_int}
F_m[Y^{(p)}_M,X^{(n)}_L; C_{X_L}]&=&\int_{-\infty}^\infty dt\, Y^{(p)}_M(t)\, {\mathcal T}_{\ln^m}[X^{(n)}_L; C_{X_L}](t)\nonumber\\
&=& (-1)^n  \int_{-\infty}^\infty  \frac{d\omega}{2\pi} (i\omega)^{n+p} \hat Y_M(-\omega) \hat X_L(\omega) A_m (\omega, C_{X_L})\nonumber\\
&=& \int_{0}^\infty  \frac{d\omega}{2\pi} (i\omega)^{n+p}[(-1)^n \hat Y_M(-\omega) \hat X_L(\omega) A_m (\omega, C_{X_L})+(-1)^p \hat Y_M(\omega) \hat X_L(-\omega) A_m (-\omega, C_{X_L})
]\,,\nonumber\\
\eea
where in the last line, as standard, we have restricted the integration domain to positive frequencies. 
Using this result the linear momentum past tails \eqref{LOtailsaver} become
\bea
\label{final_Pi_int}
(\Delta P_i)_{\rm tail} &=&  
\frac{32}{45}\pi \int_{0}^\infty \frac{d\omega}{2\pi} \omega ^{7}{\mathcal R}_i^+(\omega)
-\frac{8}{45}i\int_{0}^\infty \frac{d\omega}{2\pi} \omega ^{7}{\mathcal R}_i^-(\omega)
-\frac{2}{45}\int_{0}^\infty \frac{d\omega}{2\pi} \omega ^{8}{\mathcal S}_i^+(\omega)
+\frac{4}{63}i\pi \int_{0}^\infty \frac{d\omega}{2\pi} \omega ^{8}{\mathcal S}_i^-(\omega)
\,,\nonumber\\
(\Delta P_i)_\mathrm{tail(tail)}&=& 
-\frac{64}{45} \int_{0}^\infty \frac{d\omega}{2\pi} \omega ^{8}{\mathcal R}_i^+(\omega)\left[ 
\ln(C_{I_2} \omega  e^\gamma)\ln(C_{J_2} \omega e^\gamma)-\frac{\pi^2}{12}  +\frac{107}{105}\ln (C_{I_2}  \omega  e^\gamma)+\frac{7523}{11025}  
\right]\nonumber\\
&&
-\frac{8}{45}i\pi\int_{0}^\infty \frac{d\omega}{2\pi} \omega ^{8}{\mathcal R}_i^-(\omega)
-\frac{2}{35}\pi\int_{0}^\infty \frac{d\omega}{2\pi} \omega ^{9}{\mathcal S}_i^+(\omega)\nonumber\\
&&
-\frac{8}{63}i\int_{0}^\infty \frac{d\omega}{2\pi} \omega ^{9}{\mathcal S}_i^-(\omega)\left[  
\ln(C_{I_2} \omega  e^\gamma)\ln(C_{I_3} \omega e^\gamma)-\frac{\pi^2}{12}  +\frac{86}{105}\ln (C_{I_2}  \omega  e^\gamma)+\frac{253109}{176400}
\right]
\,,\nonumber\\
(\Delta P_i)_\mathrm{(tail)^2}  &=&
\frac{64}{45} \int_{0}^\infty \frac{d\omega}{2\pi} \omega ^{8}{\mathcal R}_i^+(\omega)\left[   
\ln(C_{I_2} \omega  e^\gamma)\ln(C_{J_2} \omega e^\gamma)+\frac{\pi^2}{4}
\right]\nonumber\\
&&
-\frac{8}{45}i\pi\int_{0}^\infty \frac{d\omega}{2\pi} \omega ^{8}{\mathcal R}_i^-(\omega)
-\frac{2}{45}\pi\int_{0}^\infty \frac{d\omega}{2\pi} \omega ^{9}{\mathcal S}_i^+(\omega)\nonumber\\
&&
+\frac{8}{63}i\int_{0}^\infty \frac{d\omega}{2\pi} \omega ^{9}{\mathcal S}_i^-(\omega)\left[   
\ln(C_{I_2} \omega  e^\gamma)\ln(C_{I_3} \omega e^\gamma)+\frac{\pi^2}{4}
\right]
\,,
\eea
\end{widetext}
where we have introduced the notation
\bea
\label{Sdefs}
{\mathcal S}_i^\pm(\omega) &=& \hat I_{ijk}(-\omega)\hat I_{jk}(\omega)\pm \hat I_{ijk}(\omega)\hat I_{jk}(-\omega)
\,,\nonumber\\
{\mathcal R}_i^\pm(\omega)&=& \epsilon_{ijk}  [\hat I_{jl}(-\omega)\hat J_{kl}(\omega)\pm \hat I_{jl}(\omega)\hat J_{kl}(-\omega)]
\,,\qquad
\eea
and
\bea
\ln(C_{I_2} \omega  e^\gamma)\ln(C_{J_2} \omega e^\gamma)=\ln^2(C_{I_2} \omega  e^\gamma)-\frac14\ln(C_{I_2} \omega e^\gamma)
\,,\nonumber\\
\ln(C_{I_2} \omega  e^\gamma)\ln(C_{I_3} \omega e^\gamma)=\ln^2(C_{I_2} \omega  e^\gamma)-\frac{7}{10}\ln(C_{I_2} \omega e^\gamma)\,.\nonumber\\
\eea

We list below the results of the computation in a large-$j$ expansion limit, referring to previous works for all technical details \cite{Bini:2021gat,Bini:2021qvf,Bini:2021jmj}.
In all cases $(\Delta P_z)_{\rm X}=0$.
The general structure is as follows:
\beq
(\Delta P_i)_{\rm tail}=\nu^2\sqrt{1-4\nu}\sum_{n=4}^{\infty}\frac{P_{i\,n}^{\rm tail}(p_\infty)}{j^n}\,,
\eeq
for dominant tails, and
\bea
(\Delta P_i)_{\rm X}&=&\nu^2\sqrt{1-4\nu}\sum_{n=5}^{\infty}\frac{1}{j^n}\left[
P_{i\,n}^{\rm X}(p_\infty)\right. \nonumber\\
&&\left.
+P_{i\,n}^{\rm X,\, {\mathcal L}}(p_\infty){\mathcal L}
+P_{i\,n}^{\rm X,\, {\mathcal L}^2}(p_\infty){\mathcal L}^2\right]\,,
\eea
with
\beq
\label{L_log}
{\mathcal L}=\ln \left(\frac{r_0 p_\infty^2}{4j}\right)\,,
\eeq
for higher-order tails $X=$ [tail(tail), (tail)$^2$].
The dependence on the scale will disappear as soon as one adds to these hereditary contributions their instantaneous counterparts (when available) at the same approximation level (see Eq. \eqref{Uvst} and the general discussion around Eq. (4.14) in Ref. \cite{Blanchet:1997jj}), giving so far a consistency check of both kinds of results.

\begin{widetext}

Tails:

\bea
(\Delta P_x)_{\rm tail}&=&\nu^2\sqrt{1-4\nu}\left[
\pi \frac{1491}{400}\frac{p_\infty^7}{j^4}
 +\frac{20608}{225}\frac{p_\infty^6}{j^5}+\pi  \frac{267583}{2400} \frac{p_\infty^5}{j^6}
+ \frac{64576}{75} \frac{p_\infty^4}{j^7}
+O\left(\frac{1}{j^8}\right)
\right]
\,,\nonumber\\
(\Delta P_y)_{\rm tail}&=& \nu^2\sqrt{1-4\nu}  \left[ 
-\frac{128}{3} \frac{p_\infty^7}{j^4}
-\pi  \frac{1509\pi^2}{140}   \frac{p_\infty^6}{j^5}+
 \left(-\frac{8768}{45} - \frac{521216\pi^2}{4725}\right)  \frac{p_\infty^5}{j^6}\right.\nonumber\\
&&\left.
+ \pi\left(\frac{36885}{896}\pi^4 - \frac{142391}{280}\pi^2\right) \frac{p_\infty^4}{j^7}
+O\left(\frac{1}{j^8}\right)
\right]
\,,
\eea

Tail of tails:

\bea
(\Delta P_x)_{\rm tail(tail)}&=&\nu^2\sqrt{1-4\nu}\left[
\frac{547712}{4725}\frac{p_\infty^9}{j^5}
+\pi^3\frac{11581}{320}\frac{p_\infty^8}{j^6}
+\left(\frac{5158208}{4725}+\frac{23902208}{55125}\pi^2\right)\frac{p_\infty^7}{j^7}
+O\left(\frac{1}{j^8}\right)
\right]
\,,\nonumber\\
(\Delta P_y)_{\rm tail(tail)}&=& \nu^2\sqrt{1-4\nu}  \left[ 
\left(\frac{29236}{245}{\mathcal L} + \frac{1509}{70}{\mathcal L}^2 + \frac{1509}{280}\pi^2 + \frac{45766481}{343000}\right)\pi\frac{p_\infty^9}{j^5}\right.\nonumber\\
&&
 + \left(-\frac{260608}{4725}\pi^2 + \frac{16678912}{1575}\ln(2)^2 + \frac{166264832}{165375}\ln(2) + \frac{8339456}{1575}{\mathcal L}\ln(2)+ \frac{1042432}{1575}{\mathcal L}^2\right.\nonumber\\
&&\left.
 + \frac{41566208}{165375}{\mathcal L} + \frac{6870773248}{5788125}\right)\frac{p_\infty^8}{j^6}\nonumber\\
&&
+ \left(\frac{80308187}{35280}{\mathcal L} + \frac{142391}{140}{\mathcal L}^2 + \frac{9664301}{896}\zeta(3) + \frac{142391}{560}\pi^2 - \frac{110655}{1792}\pi^4\right.\nonumber\\
&&\left.\left.
 + \frac{39530697589}{24696000} + \frac{110655}{32}\zeta(3){\mathcal L}\right)\pi\frac{p_\infty^7}{j^7}
+O\left(\frac{1}{j^8}\right)
\right]
\,,
\eea

Tail squared:

\bea
(\Delta P_x)_{\rm (tail)^2}&=&\nu^2\sqrt{1-4\nu}\left[
\frac{20608}{225}\frac{p_\infty^9}{j^5}
 +\frac{4569}{160}\pi^3\frac{p_\infty^8}{j^6}
+\left(\frac{299008}{875}\pi^2+\frac{64576}{75}\right)\frac{p_\infty^7}{j^7}
+O\left(\frac{1}{j^8}\right)
\right]
\,,\nonumber\\
(\Delta P_y)_{\rm (tail)^2}&=& \nu^2\sqrt{1-4\nu}  \left[ 
\left(-\frac{503}{40}\pi^2-\frac{7879}{150}-\frac{1509}{70}{\mathcal L}^2-\frac{35901}{350}{\mathcal L}\right)\pi\frac{p_\infty^9}{j^5}\right.\nonumber\\
&&
 +\left(-\frac{260608}{1575}\pi^2-\frac{16678912}{1575}\ln(2)^2+\frac{557056}{525}\ln(2)-\frac{401792}{10125}-\frac{8339456}{1575}{\mathcal L}\ln(2)\right.\nonumber\\
&&\left.\left.
-\frac{1042432}{1575}{\mathcal L}^2+\frac{139264}{525}{\mathcal L}\right)\frac{p_\infty^8}{j^6}\right.\nonumber\\
&&
+\left(\frac{853289}{960}+\frac{110655}{1792}\pi^4-\frac{142391}{240}\pi^2-\frac{1207191}{128}\zeta(3)-\frac{110655}{32}\zeta(3){\mathcal L}\right.\nonumber\\
&&\left.\left.
-\frac{142391}{140}{\mathcal L}^2-\frac{12430423}{8400}{\mathcal L}\right)\pi\frac{p_\infty^7}{j^7}
+O\left(\frac{1}{j^8}\right)
\right]
\,.
\eea
The coefficients of ${\mathcal L}^2$ in the sum $(\Delta P_y)_{\rm tail(tail)}+(\Delta P_y)_{\rm (tail)^2}$ cancel, as expected. 
We find
\bea
(\Delta P_x)_{\rm tail(tail)}+(\Delta P_x)_{\rm (tail)^2}&=& \nu^2\sqrt{1-4\nu}  \left[
\frac{196096}{945}\frac{p_\infty^9}{j^5}
 +\frac{20719}{320}\pi^3\frac{p_\infty^8}{j^6}
+\left(\frac{9226496}{4725}+\frac{42739712}{55125}\pi^2\right)\frac{p_\infty^7}{j^7}
+O\left(\frac{1}{j^8}\right)
\right]
\,,\nonumber\\
(\Delta P_y)_{\rm tail(tail)}+(\Delta P_y)_{\rm (tail)^2}&=& \nu^2\sqrt{1-4\nu}  \left[
\left(\frac{41053}{2450}{\mathcal L}+ \frac{83249503}{1029000} - \frac{503}{70}\pi^2 \right)\pi\frac{p_\infty^9}{j^5}\right.\nonumber\\
&&
+ \left(\frac{85434368}{165375}{\mathcal L} + \frac{19923246464}{17364375}-\frac{1042432}{4725}\pi^2 + \frac{341737472}{165375}\ln(2)\right)\frac{p_\infty^8}{j^6}\nonumber\\
&&\left.
+ \left(\frac{35125513}{44100}{\mathcal L} + \frac{30740778557}{12348000} -\frac{142391}{420}\pi^2 +  \frac{303491}{224}\zeta(3)\right)\pi\frac{p_\infty^7}{j^7}
+O\left(\frac{1}{j^8}\right)
\right]
\,.
\eea

\end{widetext}

\subsection{Time-symmetric tails}

In the case of time-symmetric tails one should use instead 
\bea
{\mathcal T}^{\rm ts}_{\ln^m}[X^{(n)}_L;C_{X_L}](t) &=& \int_{-\infty}^\infty \frac{d\omega}{2\pi}(-i\omega)^n\hat X_L(\omega) \nonumber\\
&\times&  
e^{-i\omega  t  }A_{m\, \rm sym}(\omega, C_{X_L})\,,
\eea
where 
\beq
A_{m\, \rm sym}(\omega, C_{X_L})=\frac12[A_m (\omega, C_{X_L})+A_m (-\omega, C_{X_L})]\,,
\eeq
with
\bea
A_{1\, \rm sym}(\omega, C_{X_L})&=&-\frac12\frac{\pi}{|\omega|}\,,\nonumber\\
A_{2\, \rm sym}(\omega, C_{X_L})&=&\frac{\pi}{|\omega|}\ln (C_{X_L}|\omega| e^\gamma)\,,
\eea
so that the typical integral is
\begin{widetext}
\bea
\label{basic_int_sym}
F_m^{\rm ts}[Y^{(p)}_M,X^{(n)}_L; C_{X_L}]&=&\int_{-\infty}^\infty dt\, Y^{(p)}_M(t)\, {\mathcal T}^{\rm ts}_{\ln^m}[X^{(n)}_L; C_{X_L}](t)\nonumber\\
&=& \int_{-\infty}^\infty \frac{d\omega}{2\pi} (-i\omega)^n (i\omega)^p \hat Y_M(-\omega)\hat X_L(\omega)A_{m\, \rm sym}(\omega)\nonumber\\
&=& \int_{0}^\infty  \frac{d\omega}{2\pi} (i\omega)^{n+p}[(-1)^n \hat Y_M(-\omega) \hat X_L(\omega) +(-1)^p \hat Y_M(\omega) \hat X_L(-\omega)]A_{m\, \rm sym}(\omega, C_{X_L})\,.
\eea

We find
\bea
\label{final_Pi_int_ts}
(\Delta P_i)_{\rm tail,\,ts}&=&  \frac{32}{45}\pi \int_{0}^\infty \frac{d\omega}{2\pi} \omega ^{7}{\mathcal R}_i^+(\omega)
+\frac{4}{63}i\pi \int_{0}^\infty \frac{d\omega}{2\pi} \omega ^{8}{\mathcal S}_i^-(\omega)
\,,\nonumber\\
(\Delta P_i)_\mathrm{tail(tail),\,ts}&=& -\frac{8}{45}i\pi\int_{0}^\infty \frac{d\omega}{2\pi} \omega ^{8}{\mathcal R}_i^-(\omega)
-\frac{2}{35}\pi\int_{0}^\infty \frac{d\omega}{2\pi} \omega ^{9}{\mathcal S}_i^+(\omega)
\,,\nonumber\\
(\Delta P_i)_\mathrm{(tail)^2,\,ts}&=& \frac{64}{45} \frac{\pi^2}{4} \int_{0}^\infty \frac{d\omega}{2\pi} \omega ^{8}{\mathcal R}_i^+(\omega)
+\frac{8}{63}i \frac{\pi^2}{4} \int_{0}^\infty \frac{d\omega}{2\pi} \omega ^{9}{\mathcal S}_i^-(\omega)
\,,
\eea
so that 
\[\begin{array}{llll}
(\Delta P_x)_{\rm tail,\,ts}&=0\,,&
(\Delta P_y)_{\rm tail,\,ts}&=(\Delta P_y)_{\rm tail}
\,,\cr
(\Delta P_x)_{\rm tail(tail),\,ts}&=(\Delta P_x)_{\rm tail(tail)}\,,\qquad&
(\Delta P_y)_{\rm tail(tail),\,ts}&=0
\,,\cr
(\Delta P_x)_{\rm (tail)^2,\,ts}&=0\,,&
(\Delta P_y)_{\rm (tail)^2,\,ts}&=(\Delta P_y)_{\rm  (tail)^2}^{\rm nolog}
\,,\cr
\end{array}\]
where \lq\lq nolog'' stands for the nonlogarithmic term of the corresponding (past tail) quantity.
In the large-$j$ expansion limit we find
\beq
(\Delta P_y)_{\rm (tail)^2,\,ts}=-\nu^2\sqrt{1-4\nu}\pi^2\left[
\frac{1509}{280}\pi\frac{p_\infty^9}{j^5}
 +\frac{260608}{1575}\frac{p_\infty^8}{j^6}
+\frac{142391}{560}\pi\frac{p_\infty^7}{j^7}
+O\left(\frac{1}{j^8}\right)
\right]\,.
\eeq

Finally, as a consequence of the above relations \eqref{LOtailsaver} and \eqref{final_Pi_int_ts} one immediately identifies the corresponding time-antisymmetric contributions. For example,
\beq
\label{final_Pi_int_ats}
(\Delta P_i)_{\rm tail,\,tas} =
-\frac{8}{45}i\int_{0}^\infty \frac{d\omega}{2\pi} \omega ^{7}{\mathcal R}_i^-(\omega)
-\frac{2}{45}\int_{0}^\infty \frac{d\omega}{2\pi} \omega ^{8}{\mathcal S}_i^+(\omega)
\,.
\eeq

\end{widetext}

Let us conclude this section by summarizing the structure of the complete expression for the energy, angular momentum and linear momentum losses, including both instantaneous and hereditary contributions. 
These radiative losses admit a double PM and PN expansion 
\bea
\frac{E^{\rm rad}}{M} &=&   \nu^2  \left[ \frac{{E}_{3}(p_\infty)}{j^3}+ \frac{{ E}_{4}(p_\infty)}{j^4}+  \cdots\right]  \,, \nonumber\\
\frac{ J_i^{\rm rad}}{J} &=&  \nu \left[ \frac{{J}_{i\,2}(p_\infty)}{j^2}+ \frac{{J}_{i\,3}(p_\infty)}{j^3} + \cdots\right]\,, \nonumber\\
\frac{P_i^{\rm rad}}{M}&=&  \nu^2\sqrt{1-4\nu} \left[ \frac{{P}_{i\,3}(p_\infty)}{j^3}+ \frac{{P}_{i\,4}(p_\infty)}{j^4}+  \cdots\right]\,,\nonumber\\
\eea
where the subscripts $n$ (e.g., in $E_n$) label  the $n$PM order, i.e., $O(G^n)$.
The subsequent expansion of the various PM coefficients in powers of $p_\infty$ then corresponds to the usual PN expansion.
Quadratic tails start at the 1.5PN fractional order, whereas cubic tails at 3PN order.
In the case of angular momentum there is an additional quadratic-in-$G$ contribution from a memory integral which is 2.5PN order.
For instance, the first few expansion coefficients of the complete (instantaneous plus hereditary) energy loss have the following structure
\bea
{ E}_{3}(p_\infty)&\sim& \underbrace{p_\infty^4}_{\rm N}+ \underbrace{p_\infty^6}_{\rm 1PN}+ \underbrace{p_\infty^8}_{\rm 2PN}+\ldots
\,,\nonumber\\
{ E}_{4}(p_\infty)&\sim& \underbrace{p_\infty^3}_{\rm N}+ \underbrace{p_\infty^5}_{\rm 1PN}+ \underbrace{p_\infty^6}_{\rm N\, tail}+ \underbrace{p_\infty^7}_{\rm 2PN}+\ldots
\,,\nonumber\\
{ E}_{5}(p_\infty)&\sim& \underbrace{p_\infty^2}_{\rm N}+ \underbrace{p_\infty^4}_{\rm 1PN}+ \underbrace{p_\infty^5}_{\rm N\, tail}+\underbrace{p_\infty^6}_{\rm 2PN}\nonumber\\
&+ & \underbrace{p_\infty^7}_{\rm 1PN\, tail}+\underbrace{p_\infty^8}_{\rm 3PN + tail(tail)+(tail)^2}+\ldots\,.\nonumber\\
\eea

A final comment concerns the dependence of the various quantities by the scale $r_0$ entering the logarithmic term ${\mathcal L}$, Eq. \eqref{L_log}.
This dependence should disappear as soon as one computes the full (gauge-invariant) linear momentum flux, i.e., including the instantaneous terms at each PN order. Consequently, the above relations will receive (and will provide too) additional confirmations from future works.

\section{The ellipticlike counterpart of linear momentum recoil hereditary effects}

Linear momentum recoil tail effects for both spinless and spinning bodies have been investigated in Ref. \cite{Racine:2008kj} at their leading PN in the simplest case of circular motion.
We compute below the corresponding past tail integrals for nonrotating bodies by including the effect of the eccentricity as series expansions in a small eccentricity parameter up to the tenth order.
We also evaluate higher-order past tail integrals (tail-of-tails and tail squared) through the same approximation level.

The Keplerian parametrization of the ellipticlike motion in harmonic coordinates is
\begin{eqnarray} \label{ellipQK2PN}
\ell&=& n t= u-e_r  \sin u\,,\nonumber\\  
r&=& a_r (1-e_r  \cos u)\,,\nonumber\\
\phi &=&2\, {\rm arctan}\left[\sqrt{\frac{1+e_r}{1-e_r}}\tan \frac{u}{2}  \right]\,,
\end{eqnarray}
with 
\beq
\label{cos_phi_sin_phi}
\cos \phi=\frac{\cos u -e_r}{1-e_r \cos u}\,,\quad \sin \phi=\frac{\sqrt{1-e_r^2}\sin u}{1-e_r \cos u}\,,
\eeq
where we have used dimensionless variables $t$ and $r$ (with $G=M=c=1$ as in the hyperboliclike case), and the orbital parameters $n=\frac{2\pi}{T_r}$ ($T_r$ denoting the radial period), $a_r$, $e_r$ are given by
\beq
n=(-2\bar E)^{3/2}\,,\quad
a_r=\frac1{(-2\bar E)}\,,\quad
e_r=\sqrt{1+2\bar Ej^2}\,,
\eeq
in terms of the dimensionless energy and angular momentum parameters $\bar E<0$ and $j$.
At the Newtonian level one can invert the relation defining the mean anomaly $\ell$ as a function of $u$ in terms of Bessel functions of the first kind as follows
\beq
u=l+\sum_{n=1}^\infty  \frac{2}{n}J_n(ne_r) \sin n l\,,
\eeq
with the property that when $l=0, 2\pi$, $u=0,2\pi$ too.
Let us mention in passing that raising the PN approximation the parametrization of the orbit includes in general three different eccentricities, $e_t$, $e_r$, and $e_\phi$, which however coincide at the Newtonian level considered here.

When dealing with the ellipticlike case instead of the integral Fourier transform one uses expansion in Fourier series, namely
each multipolar moment $X_L$ reads
\beq
X_{L}(t)=\sum_{p=-\infty}^\infty \hat X_L(p) e^{+ip\ell}\,,
\eeq    
with
\beq
\hat X_L(p)=\int_0^{2\pi} \frac{d\ell}{2\pi}\, X_L(\ell)e^{-ip\ell}\,.
\eeq

In view of the following use of Cartesian coordinates $x=r\cos\phi$, $y=r\sin \phi$ for motion on the $z=0$ plane, we also recall the useful relations
\beq
x=a_r(\cos u-e_r)\,,\qquad y=a_r\sqrt{1-e_r^2} \sin u\,,
\eeq
together with their Fourier transform counterparts, 
\bea
\hat x(p)&=&\frac{a_r}{p}J_{p}'(pe_r)\,,\nonumber\\
\hat y(p)&=&\frac{i a_r \sqrt{1-e_r^2}}{pe_r}J_p(pe_r)\,,
\eea
since
\beq
\label{BesselJ_def}
J_n(x)=\int_0^{2\pi}\frac{du}{2\pi}e^{i(n u-x\sin u)}\,,
\eeq
and
\bea
&& J_{n-1}(x)-J_{n+1}(x)=2 J'_n(x)\,,\nonumber\\
&&  J_{n-1}(x)+J_{n+1}(x)=\frac{2n}{x} J_n(x)\,,
\eea
where a prime means derivative with respect to the argument, $d/dx(J_{n}(x))|_{x=ne_r}$. 

Let us denote the average over a period of radial motion as follows
\bea
\langle (\Delta P_i)_{\rm X} \rangle &=&\frac{1}{T_r}\int_0^{T_r} dt \mathcal{F}_{P\,\rm X}^i(t) \nonumber\\
&=& \frac{1}{2\pi}\int_0^{2\pi} d\ell \mathcal{F}_{P\,\rm X}^i(\ell)\,,
\eea
with $X=$tail, tail(tail), (tail)$^2$. 
Eqs. \eqref{Ftail} and \eqref{Ftailtail} are then simply modified. 

When working in the elliptic case, instead of referring to the $x$ and $y$ components associated with nonrotating axes $\partial_x$ and $\partial_y$, it is customary to consider components with respect to the rotating axes $e_{\hat r}$ and $e_{\hat \phi}$ defined as
\bea
e_{\hat r}&=&\cos \phi(t) \partial_x+\sin \phi(t) \partial_y\,,\nonumber\\
e_{\hat \phi}&=&-\sin \phi(t) \partial_x+\cos \phi(t) \partial_y\,,
\eea
along the radial and azimuthal direction, respectively, also denoted as ${\mathbf n}=e_{\hat r}$ and ${\boldsymbol \lambda}=e_{\hat \phi}$ in the literature.
One then usually takes the following orbital averages
\bea
\langle (\Delta P_r)_{\rm X}\rangle&\equiv&
\langle \mathcal{F}_{P\,\rm X}^i(t) n_i(t) \rangle
\,, \nonumber\\
\langle (\Delta P_\phi)_{\rm X}\rangle&\equiv&
\langle \mathcal{F}_{P\,\rm X}^i(t) \lambda_i(t) \rangle
\,.
\eea
We find

\begin{widetext}

\bea
\langle (\Delta P_r)_{\rm tail}\rangle &=&-\frac{\nu^2}{a_r^7}\sqrt{1-4\nu}\left[
\frac{928}{105} {\mathbb L}-\frac{47368}{1575}+\frac{216}{7} \ln(3)-\frac{512}{35} \ln(2)\right.\nonumber\\
&&
+\left(\frac{416}{3} {\mathbb L}-\frac{20792}{45}-\frac{30402}{35} \ln(3)+\frac{42032}{21} \ln(2)\right)e_r^2\nonumber\\
&&
+\left(\frac{80056}{105} {\mathbb L}-\frac{792698}{315}+\frac{390625}{42} \ln(5)+\frac{299079}{70} \ln(3)-\frac{515008}{21} \ln(2)\right)e_r^4\nonumber\\
&&
+\left(\frac{91792}{35} {\mathbb L}-\frac{904184}{105}-\frac{660203125}{6048} \ln(5)+\frac{62965377}{1120} \ln(3)+\frac{168904676}{945} \ln(2)\right)e_r^6\nonumber\\
&&
+\left(\frac{193523}{28} {\mathbb L}-\frac{7600219}{336}+\frac{19536910589}{55296} \log (7)+\frac{181505890625}{387072} \ln(5)\right.\nonumber\\
&&\left.
-\frac{49766794011}{71680} \ln(3)-\frac{296036371}{315} \ln(2)\right)e_r^8\nonumber\\
&&
+\left(\frac{2147431}{140} {\mathbb L}-\frac{420739033}{8400}-\frac{102526991549801}{27648000} \ln(7)-\frac{554105196875}{516096} \ln(5)\right.\nonumber\\
&&\left.\left.
+\frac{21770237937411}{7168000} \ln(3)+\frac{1546844094059}{189000} \ln(2)\right)e_r^{10}
+O(e_r^{12})\right]
\,,
\eea
and
\bea
\langle (\Delta P_\phi)_{\rm tail}\rangle &=&  -\frac{\nu^2}{a_r^7}\sqrt{1-4\nu}\pi\left[
\frac{824}{35}+ \frac{1772}{5} e_r^2
+ \frac{66926}{35} e_r^4
+ \frac{32808143}{5040} e_r^6
+ \frac{3298570459}{193536} e_r^8\right.\nonumber\\
&&\left.
+ \frac{1822017899719}{48384000} e_r^{10}
+O(e_r^{12})\right]
\,,
\eea
with 
\beq
{\mathbb L}=-\frac32\ln(a_r)+\gamma+\ln(\tau_0)\,.
\eeq
In the quasi-circular case $(e_r=0)$ the previous expressions reduce to 
\bea
\langle (\Delta P_r)_{\rm tail}\rangle_{\rm circ} &=& -\frac{928}{105}\nu^2\sqrt{1-4\nu}\omega^{14/3}
\ln\left(\frac{\omega}{\omega_{\rm NS}}\right)
\,,\nonumber\\
\langle (\Delta P_\phi)_{\rm tail}\rangle_{\rm circ} &=&  -\frac{824}{35}\nu^2\sqrt{1-4\nu}\omega^{14/3}\pi
\,,
\eea
in agreement with Ref. \cite{Racine:2008kj}, where $\omega=a_r^{-3/2}$, and
\beq
 \ln \omega_{\rm NS}= -\ln(\tau_0)+\frac{5921}{1740}+\frac{48}{29}\ln(2) -\frac{405}{116}\ln(3) -\gamma\,.
\eeq

Higher-order tails turn out to be
\bea
\langle (\Delta P_r)_{\rm tail(tail)}\rangle &=&\frac{\nu^2}{a_r^{17/2}}\sqrt{1-4\nu}\pi\left[
- \frac{648}{7}\ln(3) - \frac{16}{105}\ln(2) -  \frac{1648}{35} {\mathbb L}   + \frac{1134068}{11025}
\right.\nonumber\\
&+&\left(- \frac{5588}{5} {\mathbb L} + \frac{1302389}{525} +  \frac{91206}{35}\ln(3)  -  \frac{866948}{105}\ln(2) \right) e_r^2\nonumber\\
&+&\left(- \frac{918751}{105} {\mathbb L}   + \frac{863978657}{44100} -  \frac{1953125}{42}\ln(5)   -  \frac{897237}{70}\ln(3) +  \frac{642015}{7}\ln(2) \right) e_r^4\nonumber\\
&+&
\left(- \frac{29096101}{720} {\mathbb L} + \frac{82512157631}{907200} +  \frac{3301015625}{6048}\ln(5)  -  \frac{408412611}{1120}\ln(3)  -  \frac{4409133443}{5040}\ln(2)\right) e_r^6
\nonumber\\
&+&\left(- \frac{66132493883}{483840} {\mathbb L}   + \frac{20909266973897}{67737600} -  \frac{907529453125}{387072}\ln(5) -  \frac{136758374123}{55296}\ln(7)\right.\nonumber\\
& +&\left.  \frac{300867436881}{71680}\ln(3)   +  \frac{271995624413}{53760}\ln(2)\right) e_r^8\nonumber\\
&+&
\left( - \frac{404559776017}{1075200}{\mathbb L}  + \frac{17309470577775757}{20321280000} +  \frac{717688940848607}{27648000}\ln(7)  - \frac{130779522305289}{7168000}\ln(3)\right.  \nonumber\\
&-&\left.\left.  \frac{941760801310207}{16128000}\ln(2)  + \frac{2770525984375}{516096}\ln(5)\right) e_r^{10}  +O(e_r^{12})\right]
\,,
\eea
and
\bea
\langle (\Delta P_\phi)_{\rm tail(tail)}\rangle &=& \frac{\nu^2}{a_r^{17/2}}\sqrt{1-4\nu}\left\{
\frac{2656}{21}{\mathbb L}^2+\left(-\frac{2964176}{11025}+\frac{11328}{35}\ln(2)+\frac{1296}{7}\ln(3)\right){\mathbb L}\right.\nonumber\\
&+&\frac{3488}{15}\ln(2)^2-\frac{3281552}{11025}\ln(2)-\frac{59292}{245}\ln(3)+\frac{648}{7}\ln(3)^2+\frac{1296}{7}\ln(2)\ln(3)-\frac{664}{63}\pi^2+\frac{278958328}{1157625}
\nonumber\\
&+&
\left[\frac{9104}{3}{\mathbb L}^2+\left(\frac{2047712}{105}\ln(2)-\frac{18792}{35}\ln(3)-\frac{10200152}{1575}\right){\mathbb L}
\right.\nonumber\\
&+&
\frac{3151088}{105}\ln(2)^2-\frac{2276}{9}\pi^2+\frac{2972376}{1225}\ln(3)-\frac{9396}{35}\ln(3)^2-\frac{18792}{35}\ln(2)\ln(3)+\frac{321000172}{55125}\nonumber\\
&&\left.-\frac{267930248}{11025}\ln(2) \right]e_r^2\nonumber\\
&+&
\left[\frac{502100}{21}{\mathbb L}^2+\left(-\frac{563662934}{11025}+\frac{1953125}{21}\ln(5)-\frac{353928}{7}\ln(2)
+\frac{139158}{35}\ln(3)\right){\mathbb L}\right.\nonumber\\
&&
+\frac{1953125}{21}\ln(2)\ln(5)-\frac{23828125}{196}\ln(5)+\frac{69579}{35}\ln(3)^2+\frac{17768218079}{385875}-\frac{125525}{63}\pi^2-\frac{5958117}{490}\ln(3)\nonumber\\
&&\left.+\frac{1254559802}{11025}\ln(2)-\frac{6050836}{35}\ln(2)^2+\frac{139158}{35}\ln(2)\ln(3)+\frac{1953125}{42}\ln(5)^2 \right]e_r^4\nonumber\\
&+&
\left[\frac{775974}{7}{\mathbb L}^2
+\left(-\frac{2616344179}{11025}+\frac{215372277}{280}\ln(3)+\frac{1318656788}{945}\ln(2)-\frac{1122578125}{1512}\ln(5)\right){\mathbb L}\right.\nonumber\\
&& +\frac{495336998249}{2315250}-\frac{1122578125}{1512}\ln(2)\ln(5)-\frac{12763199077}{6615}\ln(2)-\frac{1122578125}{3024}\ln(5)^2\nonumber\\
&&-\frac{19486626597}{19600}\ln(3)+\frac{7697078125}{7056}\ln(5)+\frac{215372277}{560}\ln(3)^2+\frac{434888757}{280}\ln(2)\ln(3)
\nonumber\\
&&\left.  -\frac{129329}{14}\pi^2+\frac{390500482}{189}\ln(2)^2 \right]e_r^6\nonumber\\
&+&
\left[ \frac{15788093}{42}{\mathbb L}^2+\left(\frac{550718046875}{193536}\ln(5)+\frac{136758374123}{27648}\ln(7)-\frac{35514677911}{44100}-\frac{142175867}{21}\ln(2)\right.\right. \nonumber\\
&&\left.-\frac{235903522689}{35840}\ln(3)\right){\mathbb L}+\frac{1315286367923}{132300}\ln(2)-\frac{1191751545929}{184320}\ln(7)+\frac{2242714831837}{3087000} \nonumber\\
&&
+\frac{136758374123}{55296}\ln(7)^2+\frac{46241895505311}{5017600}\ln(3)-\frac{235903522689}{71680}\ln(3)^2+\frac{136758374123}{27648}\ln(2)\ln(7)
\nonumber\\
&&-\frac{21479811151}{1890}\ln(2)^2
-\frac{23305457140625}{5419008}\ln(5)+\frac{550718046875}{193536}\ln(2)\ln(5)
+\frac{550718046875}{387072}\ln(5)^2\nonumber\\
&&\left.
-\frac{13504975371}{1024}\ln(2)\ln(3)-\frac{15788093}{504}\pi^2 \right]e_r^8\nonumber\\
&+&
\left[\frac{6220217}{6}{\mathbb L}^2
+\left(\frac{2511911887271}{23625}\ln(2)-\frac{301853979214361}{6912000}\ln(7)-\frac{358931046875}{55296}\ln(5)\right.\right.\nonumber\\
&&
\left.-\frac{559974211}{252}+\frac{6632998943769}{256000}\ln(3)\right){\mathbb L}-\frac{358931046875}{110592}\ln(5)^2+\frac{92900058064047}{1792000}\ln(2)\ln(3) \nonumber\\
&&
+\frac{176889436093}{88200}-\frac{166476155162283}{4480000}\ln(3)-\frac{1409954253851873}{9922500}\ln(2)+\frac{704817753125}{72576}\ln(5)\nonumber\\
&&+\frac{688790546856503}{11520000}\ln(7)+\frac{2221469576659}{9450}\ln(2)^2-\frac{301853979214361}{13824000}\ln(7)^2-\frac{6220217}{72}\pi^2\nonumber\\
&&\left.\left.+\frac{6632998943769}{512000}\ln(3)^2
-\frac{358931046875}{55296}\ln(2)\ln(5)-\frac{301853979214361}{6912000}\ln(2)\ln(7) \right]e_r^{10}
+O(e_r^{12})\right\}
\,,\nonumber\\
\eea
and
\bea
\langle (\Delta P_r)_{({\rm tail})^2}\rangle &=&\frac{\nu^2}{a_r^{17/2}}\sqrt{1-4\nu}\pi\left[
\frac{3296}{75}+\frac{6784}{105} \ln (2)-\frac{432}{7} \ln (3)\right.\nonumber\\
&&
+\left(\frac{6}{35} {\mathbb L}+\frac{187279}{175}+\frac{94446}{35} \ln (3)-\frac{106220}{21} \ln (2)\right)e_r^2\nonumber\\
&&
+\left(\frac{43}{24} {\mathbb L}+\frac{30585967}{3600}-\frac{5546875}{168} \ln (5)-\frac{3493611}{280} \ln (3)+\frac{5421613}{60} \ln (2)\right)e_r^4\nonumber\\
&&
+\left(\frac{9211}{1120} {\mathbb L}+\frac{11966181029}{302400}+\frac{2773890625}{6048} \ln (5)-\frac{287958243}{1120} \ln (3)-\frac{2072662519 }{3024}\ln (2)\right)e_r^6\nonumber\\
&&
+\left(\frac{1753799}{69120} {\mathbb L}+\frac{11163364495}{82944}-\frac{25013471539}{13824} \ln (7)-\frac{98475390625}{48384} \ln (5)+\frac{12206557995}{3584} \ln (3)\right.\nonumber\\
&&\left.
+\frac{347801728481}{80640} \ln (2)\right)e_r^8\nonumber\\
&&
+\left(\frac{1986762119}{32256000} {\mathbb L}+\frac{359948061095993}{967680000}+\frac{572571167150731}{27648000} \ln (7)+\frac{2428186986875}{516096} \ln (5)\right.\nonumber\\
&&\left.\left.
-\frac{15736391465067}{1024000} \ln (3)-\frac{435117633509879}{9676800} \ln (2)\right)e_r^{10}
+O(e_r^{12})\right]
\,,
\eea
and
\bea
\langle (\Delta P_\phi)_{({\rm tail})^2}\rangle &=& \frac{\nu^2}{a_r^{17/2}}\sqrt{1-4\nu}\left\{
-\frac{3088}{105} {\mathbb L}^2+{\mathbb L} \left(\frac{236704}{1575}-\frac{6176}{35} \ln (2)-\frac{432}{7} \ln (3)\right)-\frac{3088}{105} \pi^2-\frac{834812}{4725}\right.\nonumber\\
&&
-\frac{24704}{105} \ln ^2(2)-\frac{1728}{7} \ln (2) \ln (3)+\frac{792 }{7}\ln (3)+\frac{258992}{525} \ln (2)\nonumber\\
&&
+\left[-\frac{73928}{105} {\mathbb L}^2+{\mathbb L} \left(\frac{5684552}{1575}-\frac{774608}{105} \ln (2)-\frac{3132}{5} \ln (3)\right)-\frac{14782}{21} \pi^2-\frac{20083426}{4725}\right.\nonumber\\
&&\left.
+\frac{4374}{7} \ln ^2(3)-\frac{476544}{35} \ln ^2(2)-\frac{271512}{35} \ln (2) \ln (3)+\frac{17082}{5} \ln (3)+\frac{4842632}{315} \ln (2)\right]e_r^2\nonumber\\
&&
+\left[-\frac{83114}{15} {\mathbb L}^2+{\mathbb L} \left(\frac{6401426}{225}-\frac{50620}{21} \ln (2)+\frac{8289}{8} \ln (3)-\frac{5546875}{168} \ln (5)\right)-\frac{664697}{120} \pi ^2\right.\nonumber\\
&&
-\frac{45273701}{1350}-\frac{212139 }{35}\ln ^2(3)+\frac{1731136 }{21}\ln ^2(2)-\frac{421875}{28} \ln (3) \ln (5)-\frac{4609375}{28} \ln (2) \ln (5)\nonumber\\
&&\left.
+\frac{61015625}{1008} \ln (5)+\frac{9326097}{140} \ln (2) \ln (3)-\frac{3866337}{400} \ln (3)+\frac{21541046}{315} \ln (2)\right]e_r^4\nonumber\\
&&
+\left[-\frac{128363}{5} {\mathbb L}^2+{\mathbb L} \left(\frac{9896177}{75}-\frac{447468542}{945} \ln (2)-\frac{22512789}{80} \ln (3)+\frac{672921875}{3024} \ln (5)\right)\right.\nonumber\\
&&
-\frac{776091149}{30240} \pi ^2-\frac{140056279}{900}+\frac{244140625}{2016} \ln ^2(5)-\frac{35370351}{1120} \ln ^2(3)-\frac{214229248}{135} \ln ^2(2)\nonumber\\
&&
-\frac{25996875}{112} \ln (3) \ln (5)+\frac{1392240625}{1512} \ln (2) \ln (5)-\frac{4514246875}{18144} \ln (5)-\frac{45937347}{40} \ln (2) \ln (3)\nonumber\\
&&\left.
+\frac{423618219}{800} \ln (3)+\frac{10629229627}{14175} \ln (2)\right]e_r^6\nonumber\\
&&
+\left[-\frac{5221111}{60} {\mathbb L}^2+{\mathbb L} \left(\frac{805562213}{1800}+\frac{31533601}{15} \ln (2)+\frac{19666908027}{8960} \ln (3)-\frac{20763578125}{24192} \ln (5)\right.\right.\nonumber\\
&&\left.
-\frac{25013471539}{13824} \ln (7)\right)-\frac{21045404243}{241920} \pi^2-\frac{1425603047}{2700}-\frac{7373046875}{8064} \ln ^2(5)+\frac{5476831929}{4480} \ln ^2(3)\nonumber\\
&&
+\frac{2901800557}{315} \ln ^2(2)-\frac{12867859375}{13824} \ln (5) \ln (7)-\frac{71648241}{32} \ln (3) \ln (7)-\frac{3360878983}{512} \ln (2) \ln (7)\nonumber\\
&&
+\frac{275148186929}{82944} \ln (7)+\frac{670996875}{256} \ln (3) \ln (5)-\frac{159724478125}{96768} \ln (2) \ln (5)+\frac{120229540625}{145152} \ln (5)\nonumber\\
&&\left.
+\frac{169768650903}{17920} \ln (2) \ln (3)-\frac{287480861739}{89600} \ln (3)-\frac{718763033}{300} \ln (2)\right]e_r^8\nonumber\\
&&
+\left[-\frac{14394757}{60} {\mathbb L}^2+{\mathbb L} \left(\frac{2221960379}{1800}-\frac{181638989123}{4725} \ln (2)-\frac{6172985809161}{716800} \ln (3)\right.\right.\nonumber\\
&&\left.
+\frac{1600516634375}{774144} \ln (5)+\frac{211151303586083}{13824000} \ln (7)\right)-\frac{2578882161251}{10752000} \pi^2-\frac{1966592023}{1350}\nonumber\\
&&
+\frac{4747561509943}{1105920} \ln ^2(7)+\frac{4448974609375}{1548288} \ln ^2(5)-\frac{2253537635733}{286720} \ln ^2(3)-\frac{2728824825179}{23625} \ln ^2(2)\nonumber\\
&&
+\frac{302137338125}{36864} \ln (5) \ln (7)+\frac{79003303533}{4000} \ln (3) \ln (7)+\frac{183670973226937}{6912000} \ln (2) \ln (7)\nonumber\\
&&
-\frac{10106581528431983}{414720000} \ln (7)-\frac{152450810625}{14336} \ln (3) \ln (5)-\frac{3308624819375}{387072} \ln (2) \ln (5)\nonumber\\
&&
-\frac{10107015184375}{4644864} \ln (5)-\frac{102813272729619}{1792000} \ln (2) \ln (3)\nonumber\\
&&\left.\left.
+\frac{410617233459099}{35840000} \ln (3)+\frac{94171576220549}{1417500} \ln (2)\right]e_r^{10}
+O(e_r^{12})\right\}
\,.
\eea

\end{widetext}

\section{Concluding remarks}

We have computed higher-order tail (i.e., tail-of-tail and tail-squared) contributions to the linear momentum loss averaged along hyperboliclike orbits at their leading PN approximation.
Our computation uses harmonic coordinates and is conveniently performed in the frequency domain by applying techniques already developed in previous works \cite{Bini:2021gat,Bini:2021qvf,Bini:2021jmj}.
We have distinguished among past tails, time-symmetric and time-antisymmetric tails, determined by the full past interaction among the bodies, according to the proper behavior under time reversal. 
All results have been expressed as an expansion in the large angular momentum.
This work completes a previous analysis of leading-order hereditary contributions to the loss of energy, angular momentum and linear momentum along hyperboliclike orbits \cite{Bini:2021qvf}.
Due to the increasing level of accuracy of PM-based results on gravitational radiation we expect that these results will be extremely important for their low-velocity limit check. 

We have also computed quadratic and cubic past tails for ellipticlike orbits as series expansions in a small eccentricity parameter through the same level of approximation.
In this case quadratic tails were known in the quasi-circular case only, whereas cubic tails were never been explicitly computed before. 
Completing these results by the addition of all the instantaneous terms (at the level of accuracy we have computed hereditary terms here) is still an open issue.

\section*{Acknowledgments}
The authors thank  T. Damour for useful discussions at various stages during the development of the present project. DB also thanks R. Porto for valuable comments and informative mail exchanges.
DB thanks the International Center for Relativistic Astrophysics Network (ICRANet) for partial support, and acknowledges sponsorship of the Italian Gruppo Nazionale per la Fisica Matematica (GNFM) of the Istituto Nazionale di Alta Matematica (INDAM).

\end{document}